\documentclass{article}
\usepackage{epsfig,epsf}
\begin{document}
\begin{center}
{\Large\bf  Nonabelian solutions in ${\cal{N}}=4$, $D=5$  gauged supergravity} 
\vspace{0.6cm}
\\
Eugen Radu
\\
{\small
\it Department of
Mathematical Physics, National University of Ireland Maynooth, Ireland}

\vspace{0.6cm}
\end{center}
\begin{abstract}
We consider static, nonabelian
solutions  in ${\cal{N}}=4,~D=5$  Romans' gauged supergravity model.
Numerical arguments are presented
for the existence of asymptotically anti-de Sitter configurations
in the ${\cal{N}}=4^+$ version of the theory, with a dilaton
potential presenting a stationary point.
Considering the version of the theory with a Liouville dilaton potential,
we look for configurations with unusual topology. A new exact solution
is presented, and a counterterm method is proposed to compute the mass and action. 
\end{abstract}
\vspace{0.3cm}
\section{Introduction}
In the last few years, there has been increasing interest in solutions of
various supergravity models with nonabelian matter, following the discovery
by Chamseddine and Volkov  \cite{Chamseddine:1997nm} of a nontrivial
monopole-type supersymmetric vacua in
the context of the ${\cal{N}}=4$ $D=4$  Freedman-Schwarz  gauged
supergravity \cite{Freedman:1978ra}.
This is one of the few analytically known 
configurations involving both non-abelian gauge fields and gravity
(for a general review of such solutions see
\cite{Volkov:1999cc,Gal'tsov:2001tx}).
Its ten-dimensional lift was shown 
to represent 5-branes wrapped on a shrinking $S^2$ \cite{Chamseddine:1998mc}.
As discovered by Maldacena and Nu\~nez, this solution provides 
a holographic description for ${\cal{N}}=1,~D=4$ super-Yang-Mills theory
\cite{Maldacena:2001yy}.

Chamseddine and Volkov looked also for five dimensional nonabelian configurations
\cite{Chamseddine:2001hk}
in a version of ${\cal{N}}=4$ Romans' gauged 
supergravity model \cite{Romans:1985ps}
with a Liouville  dilaton potential.
The static spherically symmetric solution they found (although not in a closed form)
possesses two unbroken supersymmetries and has been shown by Maldacena and
Nastase to describe, after lifting  to ten dimensions, the
supergravity dual of an  NS5-brane wrapped on
$S^3$ with a twist that preserves only ${\cal{N}}=1$ supersymmetry in 2+1
dimensions \cite{Maldacena:2001pb}.
Both particle-like and black hole generalizations of the $D=5$ Chamseddine-Volkov 
solution are discussed in \cite{Bertoldi:2002ks} from a ten-dimensional perspective, 
where a background subtraction method to compute the mass and action of 
these configuration is also proposed.
Although the dilaton potential  $V(\phi)$
 can be viewed as an effective negative, position-dependent 
cosmological term,
these solutions do not have a simple asymptotic behaviour, 
the dilaton diverging at infinity.

However, as discussed in this paper, the situation is different 
 for the   ${\cal{N}}=4^+$ version of the
Romans  model, with a dilaton
 potential consisting of the sum of two Liouville terms. 
In this case, the dilaton field approaches asymptotically
a constant value $\phi_0$, which corresponds to an extremum of the
potential such that
$ dV/d\phi \big|_{\phi_0}=0$ and $V(\phi_0)<0$.
This makes possible the existence of
both regular and black hole solutions approaching at infinity
the anti-de Sitter (AdS) background.
Topological black holes with nonabelian hair are found as well,
in which case the three-sphere is replaced 
by a three-dimensional space  
of negative or vanishing curvature. 

Considering next the case of the Romans' model with a
 Liouville dilaton potential, 
it is natural to ask if apart from black holes discussed in \cite{Bertoldi:2002ks},
 whose horizon has spherical topology, there are also
 topological black holes with nonabelian fields.
Such solutions are known to exist in 
an Abelian truncation of the theory (see e.g. \cite{Cai:1997ii})
and also in a four dimensional Einstein-Yang-Mills (EYM) system with negative cosmological
constant \cite{VanderBij:2001ia}.
However, we find that the inclusion of nonabelian fields leads to a pathological
supergravity background:
the factor multiplying the hyperbolic or flat surface gets negative for some
finite values of the radial coordinate.
The mass and action of the spherically symmetric solutions
with a
 Liouville dilaton potential is computed 
 by using a boundary counterterm method, the standard background 
subtraction results being recovered.

A general discussion of the $D=5$ configurations admitting a translation along the
fourth spatial coordinate is presented in Section 5. 
Two new exact solutions are found by
uplifiting topologically nontrivial 
configurations of the ${\cal{N}}=4$ $D=4$  Freedman-Schwarz model.
We give our conclusions and remarks in the final section.

\section{General framework  }
\subsection{Action principle and field equations}
The bosonic matter content of the Romans' gauged supergravity 
consists of gravity, a scalar $\phi$, an
SU(2) Yang-Mills (YM) potential $A_{\mu}^I$ 
(with field strength $F_{\mu \nu}^{I}=
\partial_\mu A_\nu^I-\partial_\nu A_\mu^I+g_2 \epsilon^{IJK}A_\mu^J A_\nu^K$),  
an abelian potential
$W_{\mu}$ ($f_{\mu \nu}$ being the corresponding field strength),
 and a pair of
two-form fields.
These two form fields can
consistently be set to zero, which yields
the bosonic part of the action  
\begin{eqnarray}
\label{action5} 
I_5=\frac{1}{4 \pi}\int_{ \mathcal{M}} d^5x   \sqrt{-g}  \Big
(\frac{1}{4} \mathcal{R}
-\frac{1}{2}\partial_\mu\phi \,\partial^\mu\phi -\frac{1}{4}{\rm e}^{2a\phi}
F^{I}_{\mu\nu} F^{I \mu\nu}
-\frac{1}{4} {\rm e}^{-4a\phi} f_{\mu\nu} f^{
\mu\nu}
\\
\nonumber
- \frac{1}{4\sqrt{-g}}\epsilon^{\mu \nu \rho
\sigma\tau}F_{\mu \nu}^{I} F_{\sigma \tau}^{I}
W_{\tau}-V(\phi)\Big)
-\frac{1}{8\pi }\int_{\partial\mathcal{M}} d^{4}x\sqrt{-h}K,
\end{eqnarray}
where $a=\sqrt{2/3}$. Here 
\begin{eqnarray}
\label{dil-pot} 
V(\phi)= -\frac{g_2^2}{8} \left( {\rm e}^{-2a\phi}
+2\sqrt{2}\frac{g_1}{g_2} {\rm e}^{a\phi} \right) 
\end{eqnarray}
is the dilaton potential, $g_1$ being the U(1) gauge coupling constant. 
The last term in  (\ref{action5}) is the Hawking-Gibbons surface term, 
necessary to ensure that the Euler-Lagrange variation is well defined,
where $K$ is the trace 
of the extrinsic curvature for the boundary $\partial\mathcal{M}$ and $h$ is the
induced metric of the boundary.  

As discussed in \cite{Romans:1985ps}, the theory
has three canonical forms, corresponding to
different choices of the gauge coupling constants in 
(\ref{dil-pot}).
The case $g_2=0$ corresponds to ${\cal{N}}=4^0$ theory, where the
SU(2)$\times$U(1) symmetry is replaced by the abelian group U(1)$^4$; 
there is also a ${\cal{N}}=4^+$ version
in which $g_2=g_1\sqrt{2}$, and ${\cal{N}}=4^-$  
with  $g_2=-g_1\sqrt{2}$.
Also, for this truncation of the theory with vanishing two forms,
one can take $g_1=0$ and 
 find another distinct case.
Note that for a nonvanishing $g_2$, 
one can set its value to one,  by using a suitable rescaling of the field.

The field equations are obtained by varying the action
(\ref{action5})  with respect
to the field variables $g_{\mu \nu},A_{\mu}^I,~W_{\mu}$ and $\phi$
\begin{eqnarray}
\label{Einstein-eqs}
\nonumber
R_{\mu \nu}-\frac{1}{2}g_{\mu\nu}R &=&2~ T_{\mu \nu},
\\
\label{dil-eqs}
\nabla^2 \phi-\frac{a}{2}e^{2a\phi}F_{\mu \nu}^I F^{I\mu \nu} 
+ae^{-4a \phi} f_{\mu \nu}f^{\mu \nu}
-\frac{\partial V}{\partial \phi}&=&0,
\\
\label{YM-eqs}
\nonumber
D_{\nu}(e^{-4a \phi}f^{\mu \nu})-\frac{1}{4 \sqrt{-g}}\epsilon^{\mu \nu \rho \sigma \tau}F_{\nu
\rho}^IF_{\sigma \tau}^I&=&0,
\\
\nonumber
D_{\nu}(e^{2a \phi}F^{I\mu \nu})-\frac{1}{2 \sqrt{-g}}\epsilon^{\mu \nu \rho \sigma \tau}F_{\nu
\rho}^I f_{\sigma \tau} &=&0,
\end{eqnarray}
where the energy-momentum tensor is defined by
\begin{eqnarray}
\label{Tij}
T_{\mu \nu}&=&
\partial_{\mu} \phi \partial_{\nu} \phi
-\frac{1}{2}g_{\mu \nu} \partial_{\sigma}\phi \partial^{\sigma}\phi-
g_{\mu \nu}V(\phi)\ 
\\
\nonumber
&&+
 e^{2a \phi} 
     ( F_{\mu \rho}^I F_{  \nu \sigma}^I g^{\rho \sigma}
   -\frac{1}{4} g_{\mu \nu} F_{\rho \sigma}^I F^{I\rho \sigma})
+
 e^{-4a \phi} 
     ( f_{\mu \rho}  f_{  \nu \sigma} g^{\rho \sigma}
   -\frac{1}{4} g_{\mu \nu} f_{\rho \sigma}  f^{ \rho \sigma})
 .
\end{eqnarray}

\subsection{The ansatz}
Restricting to static solutions,
we consider a general metric ansatz on the form
\begin{eqnarray}
\label{metric-gen} 
ds^{2}= 
 A^2(r)dr^2+ B^{2}(r)d\Sigma^2_{3,k}-  C^2(r) dt^{2},
\end{eqnarray}
where $d\Sigma^2_{3,k}=d\psi^{2}+f^{2}_k(\psi) d\Omega_2^2$ denotes the line
element of a three-dimensional space with
constant curvature ($d\Omega^2=d \theta^2+\sin^2\theta d \varphi^2$ being
 the round metric of $S^2$).
The discrete parameter $k$ takes the values $1, 0$ and $-1$
and implies the form of the function $f_k(\psi)$
\begin{eqnarray}
f_k(\psi)=\left \{
\begin{array}{ll}
\sin\psi, & {\rm for}\ \ k=1 \\
\psi , & {\rm for}\ \ k=0 \\
\sinh \psi, & {\rm for}\ \ k=-1.
\end{array} \right.
\end{eqnarray}
When $k=1$, the hypersurface $\Sigma_{3,1}$ represents a 3-sphere; for $k=-1$,
it is a $3-$dimensional negative constant curvature space and it could be a
closed hypersuface with arbitrarly high genus under appropriate identifications.
For $k=0$, $\Sigma_{3,0}$ is a three-dimensional Euclidean space.

For the matter fields ansatz, we start by choosing a purely-electric
abelian field ansatz
\begin{eqnarray}
f=f_{rt}(r)dt\wedge dr,
\end{eqnarray}
the dilaton
field being  also a function only on the coordinate $r$.
 
The computation of the most general expression for $A_{\mu}^I$
compatible with the symmetries of the line-element (\ref{metric-gen})
is a straightforward generalization of the $k=1$ case 
discussed in Appendix A of \cite{Okuyama:2002mh}.
Applying the standard rules for
calculating the gauge potentials for any spacetime group
\cite{Forgacs:1980zs,Bergmann}, and taking
$A_t^I=0$ $i.e.$ no dyons, one finds the ansatz (with $\tau_a$ the Pauli spin
matrices)
\begin{eqnarray} \label{A}
A=\frac{1}{2} \Big\{ 
\tau_3(\omega(r) d \psi +\cos \theta d \varphi)
-\frac{d f_k(\psi)}{d \psi}(\tau_2 d \theta+\tau_1 \sin
\theta d \varphi) +\omega(r)f_k(\psi)(\tau_1 d \theta-\tau_2 \sin
\theta d \varphi)
 \Big\},
\end{eqnarray} 
the corresponding YM curvature being
\begin{eqnarray}
F=\frac{1}{2}\Big( \omega' \tau_3 dr\wedge d\psi + \omega' f_k\tau_1
dr\wedge d\theta - f_k \omega' \tau_2 dr\wedge d\varphi
+ (k-w^2)f_k \tau_2 d\psi \wedge d\theta  \\
\nonumber+(k-w^2)f_k \sin \theta\tau_1 d\psi \wedge d\varphi +(
w^2-k)f_k^2\sin \theta \tau_3 d\theta \wedge d\varphi
 \Big ),
\end{eqnarray}
where a prime denotes a derivative with respect to $r$. 

For a purely magnetic YM ansatz, the equation for the Abelian field
\begin{eqnarray}
\nabla_{\nu}\left ({\rm e}^{-4a\phi} f^{\nu\mu}
\right)=\frac{1}{4\sqrt{-g}}\epsilon^{\mu \nu \rho \sigma
\tau}F^{I}_{\nu\rho}F^{I}_{\sigma \tau}
\end{eqnarray}
have a total derivative structure, implying after the integration
the simple expression for $f_{tr}$
\begin{eqnarray}
\label{U1}
f_{tr}(r)=\frac{e^{4a\phi }}{\sqrt{-g}}(2 \omega^2-6k \omega+c)A^2(r)C^2(r),
\end{eqnarray}
where $c$ is a constant of integration.

\section{${\cal{N}}=4^+$ solutions}
%
The scalar potential for the ${\cal{N}}=4^+$
model $V(\phi)=- ( {\rm e}^{-2a\phi}
+2{\rm e}^{a\phi}  )/8 $
has exactly one extremum at $\phi=0$, corresponding to an 
the effective cosmological constant 
\begin{eqnarray}
\label{const}
\Lambda_{eff}=2 V(0)=-\frac{3}{4}.
\end{eqnarray}
As discussed by Romans \cite{Romans:1985ps},
the maximally symmetric AdS$_5$ spacetime is a 
solution of the theory for
 $\phi\equiv 0$ and pure gauge fields,
preserving the full ${\cal{N}}=4$ supersymmetry.

\newpage
\setlength{\unitlength}{1cm}

\begin{picture}(18,8)
\centering
\put(1,0.0){\epsfig{file=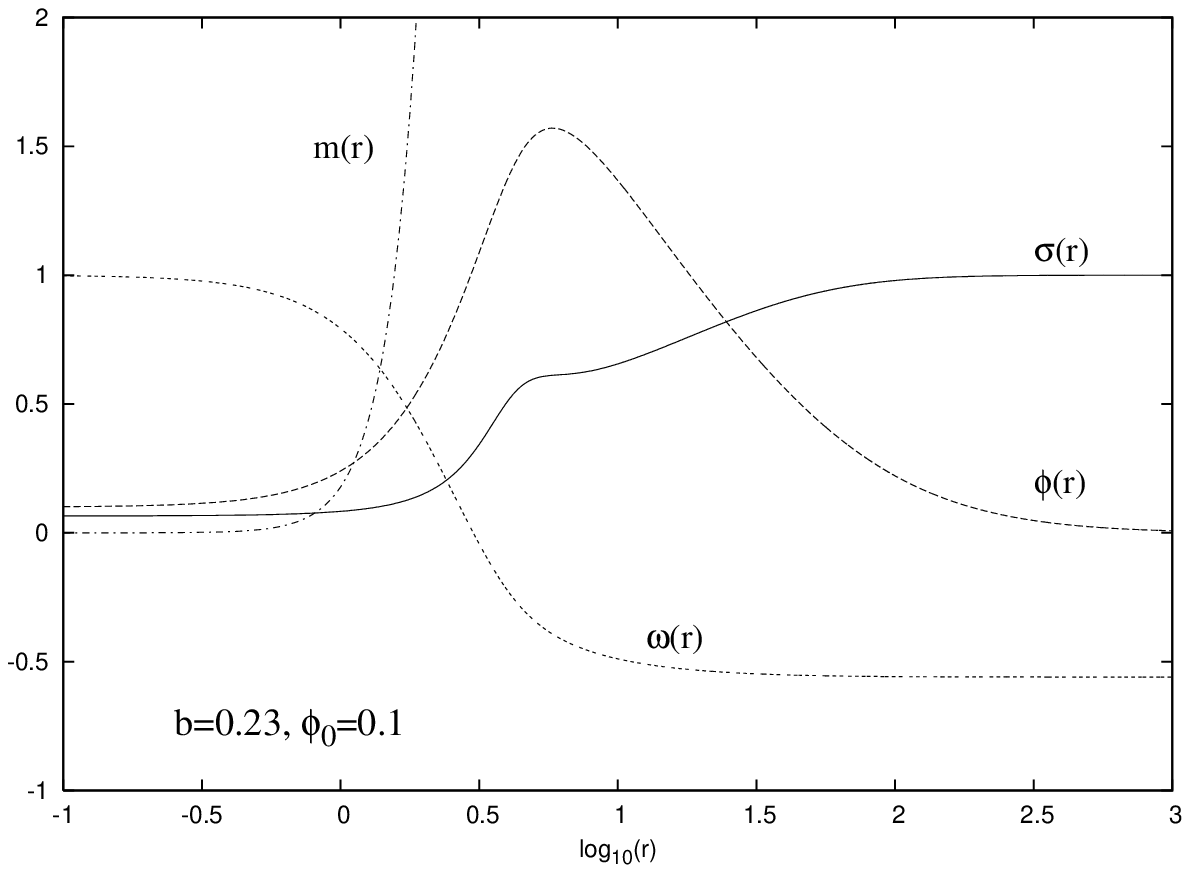,width=12cm}}
\end{picture}
\\
\\
{\small {\bf Figure 1.}
The gauge function $\omega$, the dilaton $\phi$
and the metric functions $m(r),~\sigma(r)$ are shown as function of the coordinate
$r$ for a typical spherically symmetric regular solution.
}
\vspace{0.7cm}
\\
Here we look for configurations with nonavanishing matter 
fields, approaching asymptotically the AdS background.

In deriving numerical solutions, it 
turns out to be convinient to use the following parametrization
of the line element (\ref{metric-gen})  
\begin{eqnarray}
\label{metric2}
A^2(r)=\frac{1}{H(r)},~~B^2(r)=r^2,~~C^2(r)=H(r)\sigma^2(r)
\end{eqnarray} 
The existence in the asymptotic region
 of an effective negative cosmological constant
suggest to use the following form of the metric function
$H(r)$  
\begin{eqnarray}
\label{H}
H(r)=k-\frac{4m(r)}{3r^2}+\frac{r^2}{\ell^2},
\end{eqnarray} 
(with $\Lambda_{eff}=-6/\ell^2$ $i.e.$ the characteristic length scale $\ell^2=8$), 
the function $m(r)$ being related in the usual approach
 to the local mass-energy density, up to some
numerical factor.

Inserting this ansatz into the
action (\ref{action5}), the field equations reduce to
\begin{eqnarray}
\label{eq-new}
\nonumber
m'&=&\frac{1}{2}\phi'^2H r^3+\frac{3}{2}e^{2a \phi}r(H\omega'^2+\frac{(\omega^2-k)^2}{r^2})
+\frac{1}{2}e^{4a \phi}\frac{(2\omega^3-6k\omega+c)^2}{r^3},
\\
\frac{\sigma'}{\sigma}&=&\frac{2}{3}r \phi'^2+\frac{2}{r}e^{2a \phi}\omega'^2,
\\
\nonumber
(e^{2a \phi}\sigma r H \omega')'&=&2 e^{2a \phi}\sigma \frac{\omega}{r} (\omega^2-k)
+\frac{1}{3} e^{4a \phi}\frac{6(2\omega^3-6k\omega+c)(\omega^2-k)}{r^3},
\\
\nonumber
(\sigma r^3 H\phi')'&=&3a e^{2a \phi} \sigma r (H \omega'^2
+\frac{\omega^2-k)^2}{r^2} )
+2a e^{2a \phi} \frac{\sigma}{r}\frac{(2\omega^3-6k\omega+c)^2}{r^3}
+\frac{a}{4}(e^{-2a \phi}-e^{a \phi}) r^3 \sigma,
\end{eqnarray}
while the U(1) gauge field is given by (\ref{U1}).

These equations present both globally regular and black hole
solutions.

\newpage
\setlength{\unitlength}{1cm}

\begin{picture}(18,8)
\centering
\put(1,0.0){\epsfig{file=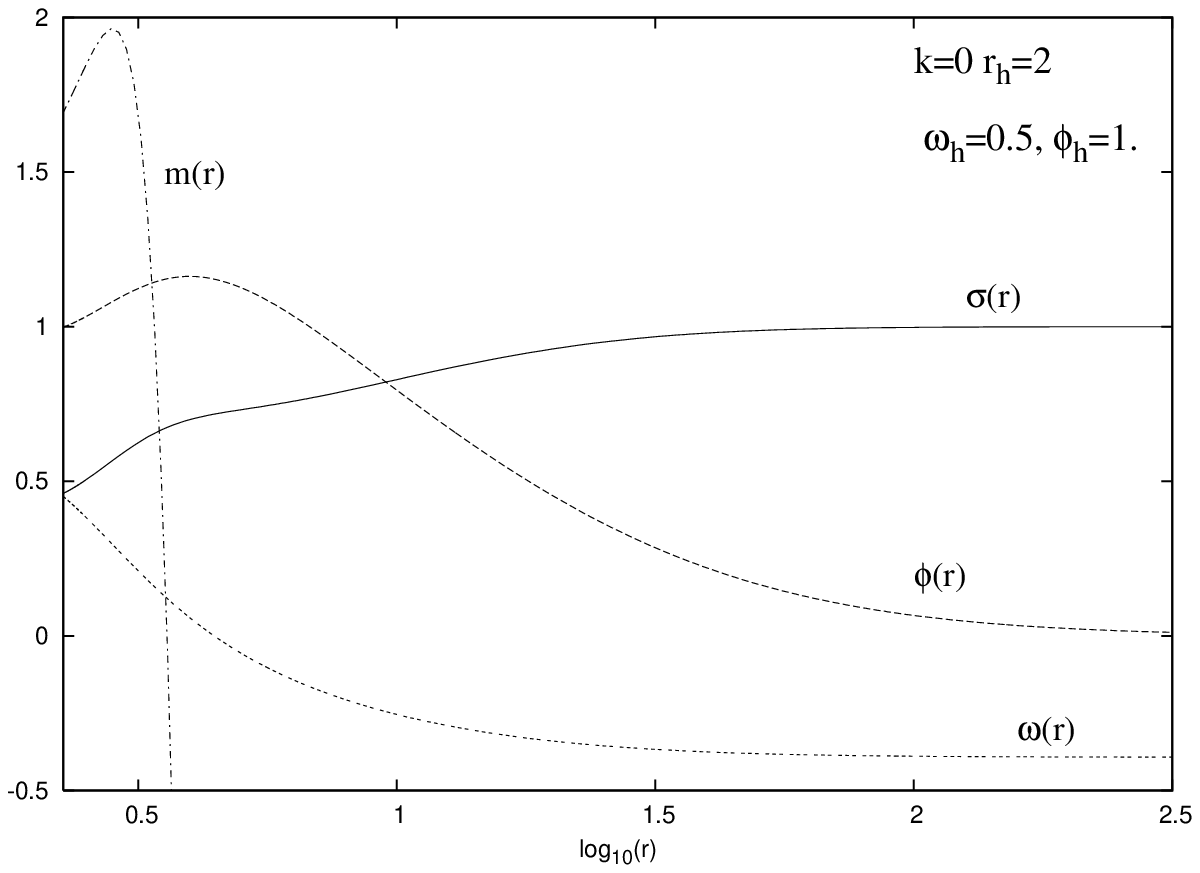,width=12cm}}
\end{picture}
\\
\\
{\small {\bf Figure 2.}
The gauge function $\omega$, the dilaton $\phi$
and the metric functions $m(r),~\sigma(r)$ are shown as function of the coordinate
$r$ for a typical  $k=0$ topological black hole solution.
}
\\ 
\\

Regular configurations exist for $k=1,~c=4$ only, and have the following expansion
 near the origin 
\begin{eqnarray}
\label{bc1}
\nonumber
m(r)&=&\big(3b^2 e^{2a\phi_0}-\frac{1}{32}( {\rm e}^{-2a\phi_0}
+2{\rm e}^{-2a\phi_0}  -3)\big)r^4+O(r^5),~~
\sigma(r)=\sigma_0(1+4b^2 e^{2a\phi_0}) r^2+O(r^2),
\\
\phi(r)&=&\phi_0+\left(3a b^2 e^{2a\phi_0}+\frac{a}{16}({\rm e}^{-2a\phi_0}
-{\rm e}^{-2a\phi_0} ) \right)r^2+O(r^4),
~~
\omega(r)=1-br^2+O(r^4),
\end{eqnarray}
with $b,~\phi_0,~\sigma_0$ real constants.

We are also interested in black hole solutions having a regular
event horizon at $r=r_h>0$.
The field equations implies
the following behaviour as $r \to r_h$ in terms of three 
parameters ($\phi_h,\sigma_h,\omega_h)$ 
\begin{eqnarray}
\label{orig}
m(r)&=& \frac{3}{4}r_h^2(k+\frac{r_h^2}{\ell^2})+m_1(r-r_h)+O(r-r_h)^2,~~
\sigma(r)=\sigma_h+\sigma_1(r-r_h)+O(r-r_h)^2,
\\
\nonumber
\phi(r)&=&\phi_h+\phi_1(r-r_h)+O(r-r_h)^2,
~~
\omega(r)=\omega_h+\omega_1(r-r_h)+O(r-r_h)^2,
\end{eqnarray} 
where we defined
\begin{eqnarray}
\label{eh}
\nonumber
m_1&=&\frac{1}{2r_h^3}\Big(e^{4a\phi_h}(2\omega_h^3-6k\omega_h+c)^2+
r_h^2(3e^{2a\phi_h}(\omega_h^2-k)^2-\frac{1}{4}r_h^4 ( {\rm e}^{-2a\phi_h}
+2{\rm e}^{-2a\phi_h} -3 ) )\Big),
\\
\omega_1&=&\frac{6r_h^2 \omega_h(\omega_h^2-k)+6e^{2a\phi_h}
(2\omega_h^3-6k\omega_h+c)(\omega_h^2-k)}
{2r_h^2(-2m_1+3kr_h+ 6r_h^3/\ell^2)},
\\
\nonumber
\phi_1&=&\frac{3(2a e^{4a\phi_h} (2\omega_h^3-6k\omega_h+c)^2+r_h^2(3a
e^{2a\phi_h} (\omega_h^2-k)^2+r_h^4V'(\phi_h) }
{2r_h^4 (-2m_1+3kr_h+ 6r_h^3/\ell^2)}.
\end{eqnarray} 
Note that since the equations (\ref{eq-new}) are invariant under
 the transformation $\omega \to -\omega$ one can set $\omega(0)>0, ~\omega(r_h)>0$ without loss
 of generality.

Using the above initial conditions,
the equations (\ref{eq-new}) were integrated
for a large set of $b,~\phi_0$ ($\omega_h,~\phi_h$ respectively) and several values of $r_h$.
For all solutions presented here we set $c=4$, although we studied 
black holes with other values of $c$ also.
The overall picture we find combines features of the
pure 
five dimensional EYM-$\Lambda$ system
discussed in \cite{Okuyama:2002mh}
and the four dimensional EYM-dilaton solutions with a potential  
approaching a constant negative value at infinity, considered in 
\cite{Radu:2004xp}. 
\newpage
\setlength{\unitlength}{1cm}

\begin{picture}(18,8)
\centering
\put(1,0.0){\epsfig{file=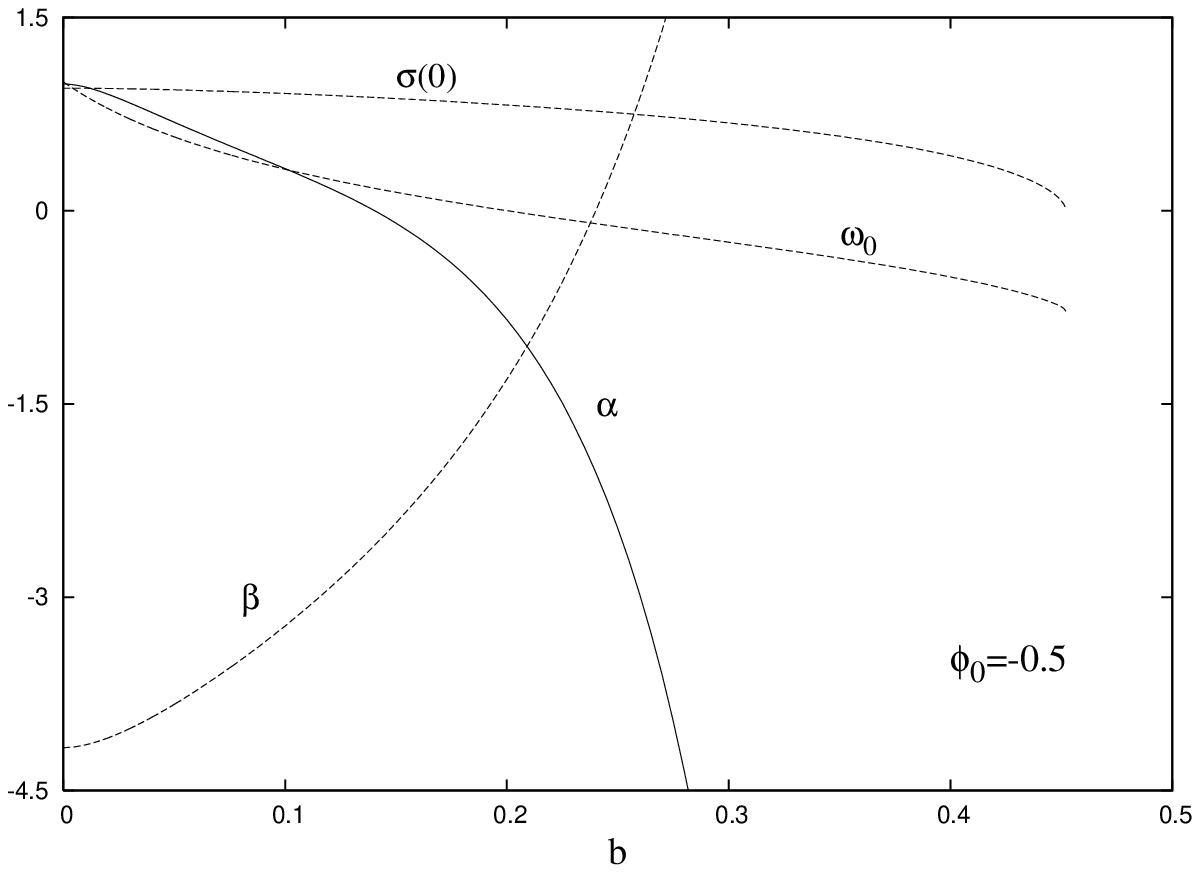,width=12cm}}
\end{picture}
\\
\\
{\small {\bf Figure 3.}
The asymptotic parameters $\omega_0,~\alpha,~\beta$,  
and the value of the metric function
function $\sigma(r)$ at the origin are shown as a function of
$b$ for spherically symmetric regular solutions with $\phi_0=-0.5$.
For $b=0$ one finds  $\alpha \simeq 0.984,~\beta \simeq -4.166$  
and $\sigma(0) \simeq 0.951$.
}
\\ 
\\
A continuum of monopole solutions 
is obtained for compact intervals of the initial
parameters.
The gauge field interpolates monotonically between the initial value at 
the $r=r_i$ (with $r_i=0$ or $r_h$) and 
some asymptotic value $\omega_0$.  
The value at the origin/event horizon 
of the  metric function $\sigma(r)$ (which is a result of the 
numerical integration)
decreases along these intervals and,
at some stage a singularity appears, corresponding to $\sigma(r_i)\to 0$.
Typical configurations are presented in Figure 1 
(a spherically symmetric regular solution) and Figure 2
(a $k=0$ black hole solution).

For all configurations we have studied,
the function $m(r)$  diverges  as $r \to \infty$.
There are two distinct sources of this behavior.
Considering first the dilaton sector, we note that
the scalar field mass $\mu$  is given by
\begin{eqnarray}
\nonumber
\mu^2=V''(0)=-\frac{1}{2}~,
\end{eqnarray}
which saturates the Breitenlohner-Freedman bound \cite{Breitenlohner:1982jf}.
Thus, the scalar field behaves asymptotically as 
\begin{eqnarray}
\label{fi-asympt}
\phi(r)=\frac{\alpha}{r^2}+\frac{\beta \log r}{r^2}+\dots,
\end{eqnarray}
where $\alpha,~\beta$ are real constants. 
For such solutions, due to the back reaction of the scalar field, 
the usual ADM mass diverges logarithmically with $r$ as $r \to \infty$
\cite{Hertog:2004dr,Henneaux:2004zi}.

There is also a second logarithmic divergence of the function $m(r)$
associated with the nonabelian gauge field $\omega(r)$.
For generic solutions, $\omega^2\to \omega_0^2\neq k$ as $r \to \infty$, 
which from (\ref{eq-new})  implies that asymptotically
$m(r) \sim(\omega_0^2-k)^2 \log r$.
A finite contribution to the total mass is found for $\omega_0^2 = k$ only.
However, we failed to find such solutions and it seems that,
similar to the EYM-$\Lambda$ case \cite{Okuyama:2002mh}, only a pure gauge 
configuration has this asymptotics. 

We remark that for any value of $\phi(r_i)$ it is
possible to find a initial value of the gauge field such that $\beta=0$,
which removes the divergencies associated with the scalar field.
Also, as seen in Figure 3 for a set of regular configurations, 
the solution with $b=0$ ($i.e.$ $\omega(r)\equiv 1$) 
and a nonzero $\phi_0$ is not the vacuum
AdS spacetime.
Thus, for these asymptotics there are globally regular 
solutions even without a nonabelian field.
One finds also black holes with scalar hair without a gauge field.
Similar solutions have been found in other models 
with a dilaton field possessing  a nontrivial
potential approaching a constant negative
value at infinity (see e.g. \cite{Radu:2004xp,Hertog:2004dr}).

A systematic analysis reveals the  following expansion of the solutions
at large $r$:
\begin{eqnarray}
\label{bc23} 
m(r)&=&M+ \frac{1}{16}\beta(\beta-4\alpha) \log r-\frac{\beta^2}{8}\log^2 r
+\frac{3}{2}(\omega_0^2-k)^2 \log r+\dots,
\\
\nonumber
\log \sigma(r)&=& -\frac{1}{3}\beta(4\alpha-\beta +2\beta \log r)\frac{\log r}{r^4}
-(\beta^2-4\alpha \beta+8\alpha^2)\frac{1}{2r^4}+\dots,
\\
\nonumber
\omega(r)&=&\omega_0-\ell^2 \omega_0(\omega_0^2-k) \frac{\log r}{r^2}+\frac{c_1}{r^2}
+\dots,
~~
\phi(r)=\frac{\alpha}{r^2}+\beta\frac{ \log r}{r^2}+\dots,
\end{eqnarray} 
where $c_1$ is a constant.
Note that the asymptotic metric still preserves the usual SO(4,2) symmetry, and the
spacetime is still asymptotically AdS, despite the diverging $m(r)$.

We close this section remarking that in the absence of a gauge field, it is
possible to define a total mass by using the approach in  \cite{Hertog:2004dr,Henneaux:2004zi}.
By employing an Hamiltonian method,
the divergencies
from the gravity and scalar parts cancel out, yielding a finite total charge.
It would be interesting to generalize this approach in the presence of a 
nonabelian gauge field in the bulk and to define a finite mass and action 
for these configurations, too.

\section{$g_1=0$ solutions}
The properties of the solutions are very different if we set $g_1=0$ in (\ref{dil-pot}),
which  leads to a Liouville-type dilaton potential  
\begin{eqnarray}
\label{dil-pot-2} 
V(\phi)= -\frac{1}{8}  {\rm e}^{-2a\phi}.  
\end{eqnarray}
In this case we found convinient to
use the following parametrization of the metric ansatz (\ref{metric-gen})
\begin{eqnarray}
\label{metric-g} 
A(r)=e^{a\phi(r)+Y(r)-X(r)},~~B(r)=e^{a\phi(r)}R(r),~~C(r)=e^{a\phi(r)+X(r)},
\end{eqnarray}
which yields 
the reduced action 
\begin{eqnarray}
\label{lagr1}
L=\frac{3}{2}e^{3a \phi +2X-Y}(R'^2+3a R\phi'R'+a R^2 \phi' X'+R R'X'+R^2
\phi'^2 +k e^{-2X+2Y})
\\
\nonumber
-\frac{3}{2}e^{3a\phi+Y}(e^{2X-2Y} R \omega'^2+\frac{\omega^2-k)^2}{R})
-\frac{1}{2}e^{3a
\phi+Y}\frac{(2\omega^3-6k\omega+c)^2}{R^3}+\frac{1}{8}e^{3a\phi+Y}R^3.
\end{eqnarray}
We remark that (\ref{lagr1}) allows for the reparametrization
$r \to \tilde{r}(r)$ which is unbroken by our ansatz.

In this way, we find that the field equations  
are (here we fix the metric gauge  by setting $Y=0$ and define $e^{2X}=\nu$)
\begin{eqnarray}               
\label{eq1} 
\nonumber
R''
=-\frac{2(2\omega^3-6k\omega+c)^2}{\nu
R^5}+\frac{2k}{\nu R}-\frac{4}{\nu R^3}(\omega^2-k)^2-\frac{\nu 'R'}{\nu}-
\frac{2 R'^2}{R}-\frac{2\omega'^2}{R}-\sqrt{6}R'\phi' ,
\\
\nonumber 
(e^{3a \phi}R \nu \omega')'=e^{3a \phi}(\omega^2-k)
\Big(
\frac{2\omega}{R}
+\frac{3}{2}\frac{(2\omega^3-6k\omega+c)}{R^3} 
\Big),
\\
\label{eqs-tot}
R''+\frac{(k-\omega^2)^2}{\nu R^3}-\frac{k}{\nu R}
+R'(\frac{\nu'}{\nu}+2\phi')+\frac{R'^2}{R}+\frac{\omega'^2}{R}=0,
\\
\nonumber  
\phi''-\frac{(k-\omega^2)^2}{\nu R^4}+\frac{k}{\nu R^2}
-\frac{\nu' R'}{\nu R}-\frac{R'^2}{R^2}-\frac{2R'\phi'}{R}=0,
\\ 
\nonumber 
X'- \alpha e^{Y-X-2s}=0,
\end{eqnarray}
where  $\alpha$ is an integration constant.
In the "extremal" case $\alpha=0$ we can set $X=0$, without loss of generality.
For $\alpha \neq 0$, we find black hole solutions with a nontrivial 
$\nu=e^{2X}$ metric function.

\subsection{BPS configurations}
As discussed in \cite{Chamseddine:2001hk}, \cite{Bertoldi:2002ks},
for  $\alpha=0,~k=1$   the system (\ref{eqs-tot}) presents
configurations preserving some amount of supersymmetry and solving first order
Bogomol'nyi  equations.

To find such solutions for any value of $k$, it is 
convinient to introduce the new variables (also, 
to conform with other results in literature, we note here $c=4\kappa$) 
\begin{eqnarray}
\phi=\sqrt{\frac{2}{3}}s-\sqrt{\frac{3}{2}}g-\sqrt{\frac{1}{6}}X,
~~~R=e^g~.
\end{eqnarray}
With this choice, the lagrangian (\ref{lagr1}) becomes
\begin{eqnarray}
\label{lagr}
L={\rm e}^{X +2s}\left(\frac{2}{3}s'^2
-\frac{1}{2}g'^2-\frac{X'^2}{3}\right)
-{\rm e}^{X +2s-2g}\omega'^2
-{\rm e}^{-X +2s-4g}(\omega^2-k)^2
\\
\nonumber
-\frac{1}{3}(2\kappa-3k\omega+\omega^3)^2 {\rm e}^{-X +2s-6g}
+k {\rm e}^{-X+Y+2s-2g}-\frac{1}{12}{\rm e}^{-X +2s},
\end{eqnarray}
which 
can be written in the form
\begin{eqnarray}
\label{lag}
L=G_{ik}(y){ dy^i\over  dr} { dy^k\over  dr} - U(y),
\end{eqnarray}
where $y^i= (s,g,w)$ and
 $G_{ik}=e^{2s } {\rm diag}(\frac{2}{3},- \frac{1}{2},-{\rm e}^{-2g})$.
The potential $U$ can be  represented as
\begin{eqnarray}
U=-G^{ik}\frac{\partial W}{\partial y^i}\frac{\partial W}{\partial y^k},
\end{eqnarray}
where the superpotential $W$  has the expression
\begin{eqnarray}
W=\frac{1}{6}e^{-g+2s}
\sqrt{
\frac{1}{2}e^{2g}-6(\omega^2-k)
+18e^{-2g}(\omega^2-k)^2-
\big(
e^{-2g}(2 \omega^3-6k\omega+4\kappa)-3\omega
\big)^2
}.
\end{eqnarray}
As a result,  we find
the first order Bogomol'nyi equations $ dy^{i}/dr=G^{ik} \partial W/\partial y^k$
\begin{eqnarray}
\nonumber
\label{BPS;eq1}
s'&=&\frac{1}{2\sqrt{2}}e^{-g+Y}\sqrt{(e^{2g}+6(2k+\omega^2)+8e^{-4g}(2\kappa-3kw+w^3
)^2+12e^{2g}(w^4-4\omega \kappa+3k^2)} ,
\\
\label{BPS;eq2}
\omega'&=&\frac{e^{-g+Y}( -e^{4g}\omega+4e^{2g}(\kappa-\omega^3)+4(k-\omega^2)
(2\kappa-3k\omega+\omega^3 )  )}{\sqrt{2}\sqrt{e^{6g}+6
e^{4g}(2k+\omega^2)+8(2\kappa-3k\omega+\omega^3)^2+12e^{2g}(3k^2-4\kappa\omega+\omega^4))}
} ,
\\
\nonumber
\label{BPS;eq3}
g'&=&\frac{e^{2g}(-e^{4g}\omega+4e^{2g}(\kappa-\omega^3)+4(k-\omega^2)(2\kappa-3k
\omega+\omega^3))}{\sqrt{2}\sqrt{(e^{4g}(2k+\omega^2)+4(2\kappa-3k\omega+\omega^3)^2+
4 e ^ { 2 g } (3k^2-4\kappa \omega+\omega^4)}} ,
\end{eqnarray}
which solve also the second-order system.
It can be proven that, after a suitable redefinition, these are the
equations derived in \cite{Volkov:1999cc} for $k=1$, by using a Killing spinor approach.
 
Unfortunately, no exact solution of these equations can be found in the general case, except
for the special values ($k=0,~\kappa=0$). For a gauge choice $Y=g$, we find the 
the new exact solution of the Romans model
\begin{eqnarray}
\label{solk=0;0}
ds^{2}=e^{2\beta
r/3+4s_0/3}(2(\beta^2-e^{-2\gamma_0(r-r_0)}))^{1/3}(-\frac{dt^2}
{2(\beta^2-e^{-2\gamma_0(r-r_0)})}+dr^2+d\Omega_{0,3}^2),
\\
\nonumber
w(r)=e^{-\beta(r-r_0)}, \ \ \ \ \
e^{2a \phi}=\frac{e^{4\beta
r/3+4s_0/3}}{(2(\beta^2-e^{-2\gamma_0(r-r_0)}))^{2/3}},
\end{eqnarray}
where  $s_0,~\beta,~\gamma_0,~r_0$ are arbitrary real constants.

A direct computation reveals that
the above line element presents a curvature
singularity  for a finite value of the radial
coordinate, $r=r^\ast$ (with $e^{-2\gamma_0(r^\ast-r_0)}=\beta^2$).
This singularity appear to be repulsive: no timelike geodesic hits it,
though a radial null geodesic can.
Thus our solution violate the criterion of \cite{Maldacena:2001mw}
because $g_{tt}$ in the Einstein frame is
unbounded at the singularity and thus they cannot accurately
describe the IR dynamics of a dual gauge theory.

Of course, in other cases, the equations (\ref{BPS;eq2}) can be solved numerically.
However, a direct inspection of the above relations reveal that 
the absence
of solutions with a regular origin for $k\neq 1$ is a generic property of
$k=0,~-1$ BPS solutions (we call regular origin the point $r=r_0$,
 where the function $R(r)$ vanishes but all
curvature invariants are bounded (without loss of generality we can set $r_0$=0)).
 
It can be proven that this is a generic feature of all $k \neq 1$ configurations.
Considering solutions of the second order equations (\ref{eqs-tot}),
one finds that it is also not possible to take a consistent set 
of boundary 
conditions at the origin 
 without introducing a curvature singularity.  
This fact has to be attributed to the particular form of the potential term
$V_{YM}=e^{2a \phi} (\omega^2-k)^2/(2R)$ in the reduced lagrangean of the system.

Solutions with regular origin may exist for $k=1,~\kappa=1$ only and are parametrized 
by the value of the parameter $b$ appearing in the expansion of $\omega(r)$ at the origin
$\omega(r)=1-br^2+O(r^4)$. 
As discussed in \cite{Bertoldi:2002ks}, globally regular
solutions extending to infinity ($i.e.$ an unbounded $R(r)$) exists for $0< b <1$ only, 
the BPS solution found in \cite{Chamseddine:2001hk}
corresponding to $b=1/3$. 
For $b\geq 1$, the function $R(r)$ goes to zero again at some
finite value of $r$. 

\subsection{Black hole solutions}

A natural way to deal with the type of singularities we have found for $k=0,-1$ 
is to hide them inside an
event horizon.
To implement the black hole interpretation we restrict the parameters 
so that the metric describes the exterior of a black hole with a non-degenerate
horizon.
That implies the existence
of a point $r=r_h$ where $e^{2X}=\nu$ vanishes, while all other functions
are finite and differentiable.
Without loss of generality we can set $r_h=0$.


The field equations (\ref{eqs-tot}) give the following expansion near the event horizon
\begin{eqnarray}
\nonumber
R(r)&=&R_h+R_1r+O(r^2), ~~\phi(r)=\phi_h
+\phi_1 r+O(r^2),~~\nu(r)= \alpha \frac{e^{-3a \phi_h}}{R_h^3}r+O(r^2),
\\
\label{expansion-eh}
\omega(r)&=&w_h+\frac{2e^{3a\phi_h}}{3 \alpha R_h}(\omega^2-k)(6R_h^2
\omega_h +2\omega_h^3-6k\omega_h+c)r+O(r^2), 
\end{eqnarray}
where
\begin{eqnarray}
\nonumber
\label{expansion2}
\phi_1=\frac{a e^{3a\phi_h}}{8\alpha R_h^3}
\Big(
R_h^6-12kR_h^4+36R_h^2(\omega^2-k)^2+20(2\omega_h^3-6k\omega_h+c) 
\Big),
\\
\nonumber
R_1=\frac{e^{3a\phi_h}}{2\alpha R_h^2}
\Big(
R_h^2(k R_h^2+ (\omega_h^2-k)^2+(2\omega_h^3-6k\omega_h+c)^2 -12
e^{-3a\phi_h}\alpha a \phi_1 R_h
\Big).
\end{eqnarray}
The solutions present three free parameters:
the value of the dilaton at the horizon $\phi_h$,
the event horizon radius $R_h$ and the value of the gauge potential at the horizon
$\omega_h$.
For any $k$, one can set $\alpha=1$ without loss of generality,
since this value can be obtained by a global rescaling of the line element (\ref{metric-gen}).

Using the initial conditions on the event horizon
(\ref{expansion-eh}), the equations (\ref{eqs-tot})  were integrated
for a range of values of $\phi_h,~R_h$
and varying $\omega _{h}$.
The numerical analysis shows the existence of a continuum of solutions
for every value of ($k,R_h, \omega_h, \phi_h)$.
Also, for every choice of $\phi_h$ and a given ($k,R_h,\omega_h)$, we find qualitatively
similar solutions
(different values of $\phi_h$ lead
to global rescalings of the solutions).

The results we found for $k=1$ are similar to those derived in \cite{Bertoldi:2002ks}
from a ten-dimensional point of view.
One can show that in this case $\omega_h$ is restricted to $|\omega_h|\leq 1$, 
while the value of $c$ is not restricted. 
Spherically symmetric black hole solutions are found for any set of 
values $(\phi_h,R_h,\omega_h)$.
For $R_h^2+\omega_h^2>1$, $R(r)\to\infty$ as $r\to\infty$, while for 
$R_h^2+\omega_h^2<1$, as numerically found in
\cite{Bertoldi:2002ks}, the asymptotic is different:
$R(r)$ vanishes at some finite value of $r$, where there is a curvature singularity.

Unfortunately, the situation for a nonspherically symmetric event horizon resembles
this last case.
For each set  ($R_h,~w_h,~\phi_h$) we find
a solution living in the interval $r\in[0,r^\ast[$, where
$r^\ast$ has a finite value.
For fixed $R_h,~\phi_h$,
the value of $r^\ast$
decreases when increasing $\omega _{h}$.
A typical $k=-1$ solution is presented in Figure 4 for $c=1,~\alpha=1$.
\newpage
\setlength{\unitlength}{1cm}

\begin{picture}(18,8)
\centering
\put(1,0.0){\epsfig{file=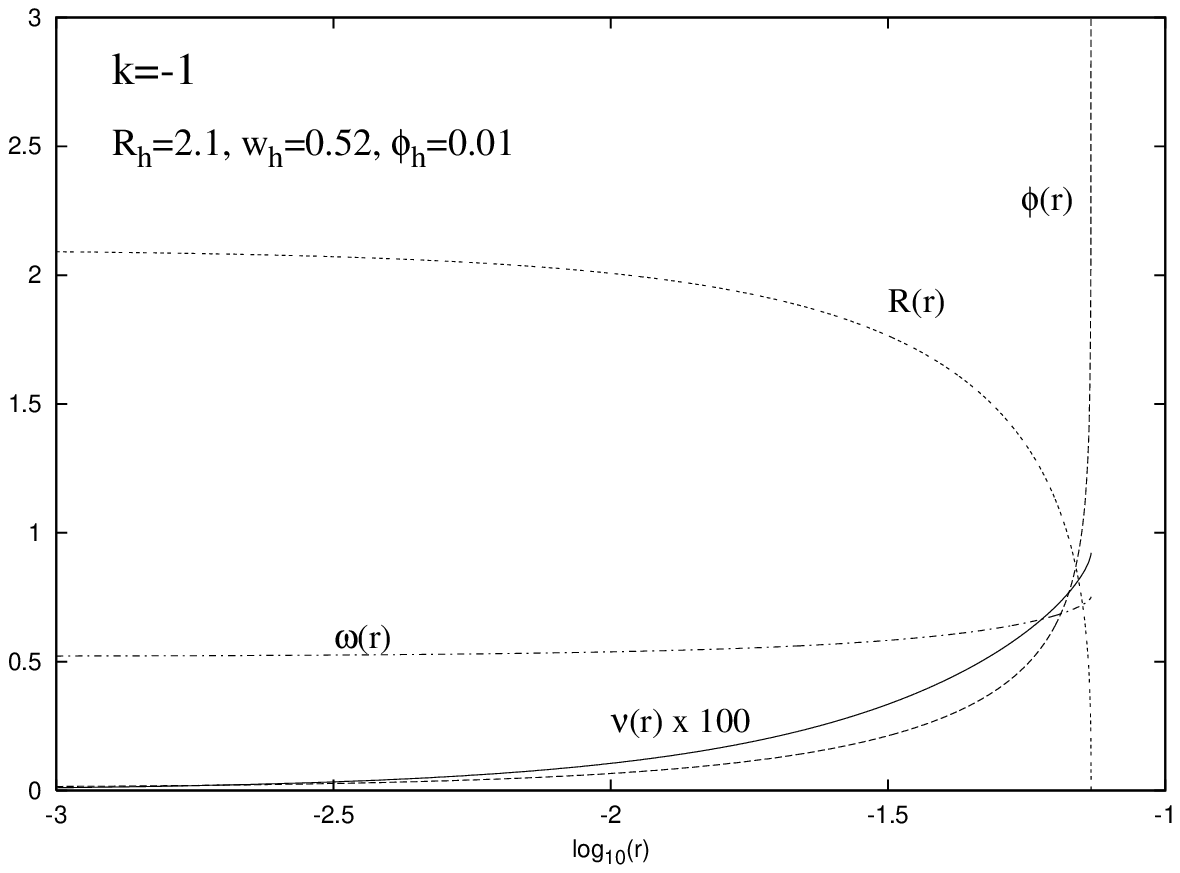,width=12cm}}
\end{picture}
\\
\\
{\small {\bf Figure 4.}
 The gauge function $\omega$, the dilaton $\phi$
and the metric functions $R,~\nu=e^{2X}$ are shown as function of the coordinate
$r$ for a typical  $k=-1$ solution with an event horizon at $R_h=2.1$.
}
\\
\\
This fact can be understood by noticing that, for $k \neq 1$, 
the boundary conditions (\ref{expansion-eh})
imply that $R(r)$ is a strictly decreasing function near the event horizon ($i.e.$ $R'(r_h)<0$).
In all cases we studied, $R(r)$ 
keeps decreasing for increasing $r$, vanishing  at $r^\ast$
where there is a curvature singularity.
For $r>r^\ast$, a wrong signature spacetime is found.

Therefore, it seems that, for a Liouville dilaton potential,
 all nonabelian configurations with nonspherical
topology present some pathological features.
Given the presence of the naked singularities,
the physical significance of these solutions is not obvious.
A similar property has been noticed in \cite{Radu:2002za} for $k=0,-1$ solutions of 
${\cal{N}}=4$ $D=4$  Freedman-Schwarz gauged supergravity model 
which possesses also a Liouville
dilaton potential.

\subsection{A counterterm proposal}
In this section we'll concentrate on the computation of mass and action of the 
generic $k=1$ spherically symmetric 
solutions for which $R(r) \to \infty$ as  $r\to \infty$.
We start by presenting the asymptotic expression  of these solution \cite{Bertoldi:2002ks},
which is shared by both globally regular and black hole configurations
\begin{eqnarray}
\label{asympt}
R &=& \sqrt{2 r} - \left(\frac{\gamma^2}{4 \sqrt{2} r^{3/2}} + 
\ldots \right) + \sqrt{2} {\cal P} r^{3/4} e^{-2r} (1+ \frac{1}{r}  +\ldots)
+ {\cal O}(e^{-3r})\,,  
\\
\nonumber
\phi &=& \frac{\phi_{\infty}}{\sqrt{6}}+\sqrt{\frac{2}{3}} r -\frac{1}{4}\sqrt
{\frac{3}{2}}\log r 
+\left(\frac{5\sqrt{3} \gamma^2}{64 \sqrt{2} r^2} + \ldots
\right)
+ \sqrt{\frac{3}{2}}{\cal P} r^{1/4}e^{-2r}(1 + \frac{1}{2r}  + \ldots) +
{\cal O}(e^{-3r})\,, 
\\
\nonumber
\omega &=& \frac{\gamma}{\sqrt{r}}\left(1 + \ldots \right)
+ {\cal C} r^{1/2} e^{-2r}(1 + \ldots) + {\cal O}(e^{-3r}),
~~~
\nu = 1 -\frac{K}{2^{5/2}\, r^{3/4}} e^{-2r + a\phi_{\infty} }(1+....),
\end{eqnarray}
where $\gamma,~K,~{\cal C},~\phi_{\infty}$ and ${\cal P}$ are free parameters.

The construction of the conserved quantities for this type of asymptotics
is an interesting problem.
We start by evaluating the on-shell action.
By using the scalar and abelian field equations, one can show
the volume term in the action (\ref{action5}) reduces to the
integral of a total derivative
\begin{eqnarray}
\label{i2}
I_{B}=\frac{1}{4 \pi}\int_{ \mathcal{M}} d^5x   \sqrt{-g} 
\left( 
 -\frac{1}{3a}\nabla^2 \phi -\partial_{\mu}(\sqrt{-g}e^{-4a\phi}f^{\mu
\nu}W_{\nu})
\right),
\end{eqnarray}
and thus can be expressed in terms of surface integrals.
However, the contribution of the abelian gauge field in the above expression can easily
be proven to vanish.
As a result, considering the metric ansatz (\ref{metric-g}) and Wick rotating 
$t \to i\tau$, we find the following expression of
the total Euclidean  action (where we included also the Hawking-Gibbons boundary term)
\begin{eqnarray}
\label{itot}
I=\frac{1}{4 \pi}\beta V_3 
\lim_{r \to \infty} \left( 
e^{2X}(e^{3a\phi}R^3)'+\frac{1}{2}R^3 e^{3 a \phi} (e^{2X})'
\right)
=\frac{1}{8 \pi}\beta V_3 \alpha+\frac{1}{4 \pi}\beta V_3 
\lim_{r \to \infty} \ 
e^{2X}(e^{3a\phi}R^3)',
\end{eqnarray}
where $V_3$ is the  three-sphere area
(note that we used the 
equation of motion for $X$ in the derivation of the last term). 
Here $\beta$ is the periodicity of the time coordinate on the Euclidean section.
For regular solutions, $\beta$ takes arbitrary values (while $\alpha=0$). 
The value of $\beta$  for black hole solutions is 
fixed by requiring the absence of conical singularities
\begin{eqnarray}
 \beta=\frac{4 \pi}{\alpha }e^{3a \phi_h}R_h^3.
\end{eqnarray} 
As expected,  the total action
presents an infrared divergence,
as implied by the asymptotic expressions (\ref{asympt}).
This divergence is associated with the infinite volume of the spacetime manifold. 

A common approach - background subtraction, uses a second, reference
spacetime to identify
divergences which should be subtracted from the action. 
After subtracting the (divergent) action of the reference background, 
the resulting action will be finite.
This is the method used in previous
studies on gauged supergravities with 
a Liouville-type dilaton potential and noanbelian fields   
\cite{Gubser:2001eg}, \cite{Bertoldi:2002ks}. 
In these cases, reference spacetime was taken to be the corresponding nonabelian BPS solution.
As found in \cite{Bertoldi:2002ks}, the mass and action  of the  non-BPS solutions
computed in 
this way
generically  diverges.
However, among all black hole and regular solutions, 
there are special configurations with finite mass
which form a discrete set, corresponding to $\gamma=0$ in the asymptotic
expansion (\ref{asympt}).

At a conceptual level, the background subtraction method is not
entirely satisfactorly, since it relies on
the introduction of a spacetime which is auxiliarly to the problem.
In some cases the choice of reference spacetime is ambiguous  -for example
NUT-charged solutions (see e.g. the discussion in \cite{Mann:2002fg}). 
This method requires also a complicated matching procedure of some
matter fields on the boundary.
It would be
nice to have a method that is intrinsic to the solutions at hand, instead of
one which requires another solution to compare to.

For asymptotically AdS spacetimes, this
problem  is solved by adding additional surface terms to the theory action 
\cite{Balasubramanian:1999re}.
These counterterms are built up with
curvature invariants of a boundary $\partial \cal{M}$ (which is sent to
infinity after the integration)
and thus obviously they do not alter the bulk equations of motion.
This yields a finite action and mass of the system.
The generalization to asymptotically flat case was
considered recently in \cite{Mann:2005yr} (see also \cite{Kraus:1999di}).
The boundary counterterm approach
in the case of a Lioville-type dilaton potential is discussed in \cite{Cai:1999xg},
however by assuming a power series decay at infinity 
which does not cover the asymptotics (\ref{asympt}).

For solution with an asymptotics given by (\ref{asympt}), we find that by adding to
the total Lorentzian action (\ref{action5}) an AdS-like boundary couterterm of the form
\begin{eqnarray}
\label{Ict}
I_{ct}=
-\frac{1}{8\pi}
\int_{\partial M}d^{4}x\sqrt{-h }
(\frac{2}{\ell}+\frac{\ell
\mathcal{R}}{8}),~~{\rm~where}~~\ell=\sqrt{-8V(\phi)}=e^{-a\phi},
\end{eqnarray}
the divergence disappears for $\gamma=0$ solutions, and we arrive at the simple
finite expression for the total Euclidean action $I{=}I_{\rm bulk}{+}I_{\rm surf}{+}I_{\rm ct}$
\begin{eqnarray}
\label{itot1}
I=\frac{1}{8\pi}\beta
V_3\left(\ 
3\sqrt{2} e^{\phi_\infty}{\cal P}-\frac{1}{2}\alpha
\right).
\end{eqnarray}
Using these counterterms one can also
construct a divergence-free stress tensor  by defining 
\begin{eqnarray}
\label{s1}
 T_{ab}=\frac{2}{\sqrt{-h}} \frac{\delta I}{ \delta h^{ab}}
=\frac{1}{8\pi }(K_{ab}-Kh_{ab}-\frac{2}{\ell}h_{ab}+\frac{1}{2}\ell E_{ab}),
\end{eqnarray}
where $E_{ab}$ is the Einstein tensor of the intrinsic metric $h_{ab}$. The
conserved charge associated with time translation is the mass of 
spacetime, which for $\gamma=0$ is given by
\footnote{Note that the action and mass of the generic solutions 
diverges like $e^{c_1r}{r^{-c_2}}\gamma^{c_3}$, with $c_i$ positive constants depending on the
free parameters which enter (\ref{asympt}).}
\begin{eqnarray}
\label{mass}
M=\int_{\partial \Sigma}d^{3}x\sqrt{-h}T_t^t=
\frac{1}{8\pi}3\sqrt{2}V_3 \beta e^{\phi_\infty}{\cal P},
\end{eqnarray}
which agrees with the results in \cite{Bertoldi:2002ks}, derived  
by using the background subtraction approach.

Once we have the renormalized action, standard techniques allows us to calculate
the full thermodynamics of the black hole.
It can be proven that the entropy of these black holes is one quarter of
the event horizon area, as expected.
By considering
the class of regular stationary metrics forming an ensemble of
thermodynamic systems at equilibrium temperature
$T=\beta^{-1}$ (see e.g. \cite{Mann:2002fg}), and applying the standard formalism
one finds
\begin{eqnarray}
\label{S}
S=\beta M-I=\frac{V_3}{16 \pi}\alpha \beta=\frac{1}{4}A_h.
\end{eqnarray} 

\section{Four dimensional reduction}
Apart from Schawarzschild-Tangherlini  connfiguration, 
the five dimensional Einstein gravity
present also other classes of solutions.
The simplest one consist in the direct product of Schwarzschild black hole and 
one extra dimension, describing an uniform black string.

It would be interesting to find the corresponding configurations 
in the ${\cal{N}}=4$ Romans model.
In this context, we consider first the  
Kaluza-Klein reduction of the action 
principle (\ref{action5}) with respect the Killing vector
$\partial/\partial x^5$.
While the
 five-dimensional metric has the usual parametrization
\begin{eqnarray}
\label{metricaKK}
ds^2 = e^{- 2\psi/\sqrt{3}}\gamma_{ij} dx^{i}dx^{j}
 + e^{ 4\psi/\sqrt{3}}(dx^5 + 2J_{i}dx^{j})^2,
\end{eqnarray}
(with $i,j=1,\dots 4$),
to reduce the gauge fields it is convinient to use the ansatz
\begin{eqnarray}
\label{ans1}
W=u(dx^5+2J_idx^i)+{\cal W}_{i}dx^i,~~
A=\Phi(dx^5+2J_idx^i)+{\cal A}_{i}dx^i,
\end{eqnarray}
where $u$ and $\Phi$ are four dimensional scalars; ${\cal W}_{i},~{\cal A}_{i}$  are gauge
fields in $D=4$ with the coresponding field strength tensors
${\cal G}_{ij}$ and ${\cal F}_{ij}$ respectively. 
This ansatz ensures that ${\cal W}_{i}$ and ${\cal A}_{i}$ 
are invariant under the coordinate transformation
$x^5 \to x^5 +f(x^i)$.
The four dimensional action takes a relatively simple form written in terms of the modified 
field strength
\begin{eqnarray}
\label{ans2}
{\cal G}_{ij}'={\cal G}_{ij}+2u M_{ij},~~{\cal F}_{ij}'={\cal F}_{ij}+2\Phi M_{ij},
\end{eqnarray}
where $M_{ij}=\partial_iJ_j-\partial_jJ_i$.

After Kaluza-Klein reduction, the bulk term in (\ref{action5})
leads to the four dimensional action principle 
\begin{eqnarray}
\label{action4}
I_4=\frac{1}{4\pi}\int d^{4}x\sqrt{-\gamma }\Big[ 
\frac{\mathcal{R} }{4}
-\frac{1}{2}\nabla_{i}\psi \nabla^{i}\psi
-\frac{1}{2}\nabla_{i}\phi \nabla^{i}\phi
-e^{2\sqrt{3}\psi}\frac{1}{4}M_{ij }M^{ij } 
-e^{2\psi/\sqrt{3}+2a \phi}\frac{1}{4} 
 {\cal F}^{'I}_{ij}  {\cal F} ^{'I ij }
\\
\nonumber
 -e^{-4\psi/\sqrt{3}+2a \phi}\frac{1}{4} D_i\Phi^I D^i\Phi^I
 -e^{2\psi/\sqrt{3}-4a \phi}\frac{1}{4} 
 {\cal G}^{'}_{ij}  {\cal G} ^{'ij } 
 -e^{-4\psi/\sqrt{3}-4a \phi}\frac{1}{4} 
\nabla_{i}u \nabla^{i}u
\\
\nonumber 
-V(\phi)e^{-2\psi/\sqrt{3}}
-\frac{u}{\sqrt{-\gamma }} \epsilon^{ijkl}{\cal F} ^{'I}_{ij }{\cal F} ^{'I}_{kl}
-\frac{4}{\sqrt{-\gamma }} \epsilon^{ijkl}{\cal W}_{i}{\cal F} ^{'I}_{jk }D_l\Phi^I
\Big],
\end{eqnarray}
which describes a system with an SU(2) gauge field, two U(1) fields and three scalars
coupled to gravity. 
We expect  the existence of monopole and dyon solutions within the above action
principle, generalizing the configurations considered in 
\cite{Volkov:2001tb}, \cite{Brihaye:2004kh}, \cite{Brihaye:2005pz}.  

The picture simplifies for a five dimensional ansatz with $A_5=0,~J_i=0$ and a Liouville
potential.
In this case,  we can consistently set to zero the $D=5$ abelian field $W_\mu$,
and the four dimensional system (\ref{action4})
admits a global symmetry $\psi \to \psi-\sqrt{2} \epsilon, ~~\phi\to \phi+\epsilon$ 
\cite{Chamseddine:2001hk}.
As a result, the following condition
 \begin{eqnarray}
 \label{cond}
\psi=\frac{1}{\sqrt{2}} \phi,
\end{eqnarray}
can be imposed and we end up with a consistent truncation of the ${\cal{N}}=4$ $D=4$ gauged 
supergravity action \cite{Freedman:1978ra}
\begin{eqnarray}
\label{actionFS}
I_4=\frac{1}{4\pi}\int d^{4}x\sqrt{-\gamma }\Big[ 
\frac{\mathcal{R} }{4}
-\frac{1}{2}\nabla_{i}\phi \nabla^{i}\phi 
-e^{2\phi }\frac{1}{4} 
 F_{ij }^I  F ^{I ij }
+\frac{1}{8}e^{-2 \phi }
\Big].
\end{eqnarray}
We can use this result to find new solutions 
of the Romans' gauged supergravity with $g_1=0$ by uplifting
known  solutions of the Freedman-Schwarz ${\cal{N}}=4$ $D=4$ 
model.  
The general $D=4$ regular solutions (generically non-BPS) will describe
$D=5$ nonabelian vortices; there exist also nonabelian black strings, obtained by
uplifting the $D=4$ black hole solutions discussed in \cite{Gubser:2001eg}.
An interesting case is provided by the topological 
BPS solutions found in \cite{Radu:2002za}.
Together with the $k=1$ exact solution presented in \cite{Chamseddine:1997nm}, this gives
a class of 
five-dimensional vortex-type solutions with 
\begin{eqnarray}
\label{k=0}
ds^2=e^{4\phi/3}\big(dr^2+R^2(r)(d\theta^2+f_k^2(\theta)
d\varphi^2)-dt^2\big)+e^{4\phi/3}(dx^5)^2,
\\
\nonumber
W_{\mu}=0,~~A=\frac{1}{2} \Big(
  \omega(r) \tau_1   d \theta
+\big(\frac{df_k(\theta)}{d\theta} \tau_3
+ f_k(\theta) \omega(r) \tau_2   \big)   d \varphi \Big)
\end{eqnarray}
and 
\begin{eqnarray}\nonumber
&&k=1:~~~\omega(r)=\frac{r}{\sinh r},~~R^2(r)=2 r \coth r-\omega^2(r)-1,
~~~e^{2(\phi(r)-\phi_0)}= \frac{\sinh r}{2 R(r)},
\\
&&k=0:~~~\omega(r)=e^{-r},~~~
R^2(r)=c-\omega^2(r),~~~e^{2(\phi(r)-\phi_0)}=\frac{e^{r}}{R(r)},
\\
\nonumber
&&k=-1:~~\omega(r)=\frac{r }{\sinh(r+c)},~~
R^2(r)=2r\coth(r+c)-\omega^2(r)-1,~~
e^{2(\phi(r)-\phi_0) }=\frac{\sinh (r+c)}{R(r)},
\end{eqnarray}
where $c$ and $\phi_0$ are arbitrary constants.
However, we see that the pathological features of the $k\neq 1$ solutions found in
$D=4$ persist after uplifiting them to $D=5$ and they still presents a naked
singularity for some finite value of the radial coordinate.

We close this Section by remarking that new nontrivial $D=4$ nonabelian solutions can
be obtained by adjusting to this case the boosting
procedure proposed in \cite{Brihaye:2005pz}.
In this approach one starts with a purely magnetic static configuration of the 
 Freedman-Schwarz ${\cal{N}}=4$ $D=4$ 
model (\ref{actionFS}),
 and uplift it  according to (\ref{metricaKK}), (\ref{ans1}) (there we take $J_i=W=\Phi=0$),
finding
in this way a vortex-type (or black string) solution of the 
$D=5$ Romans' theory.
The next step is to boost the $D=5$ solution in the 
 $(x^5,~t)$ plane
$x^5=\cosh \beta~z+\sinh \beta~ \tau,~~t=\sinh \beta~z+\cosh \beta~\tau.$
The dimensional reduction of the boosted five dimensional  configurations
along the $z-$direction provides new nontrivial $D=4$
solutions.
However, these configurations do not satisfy the
field equations of the Freedman-Schwarz model, 
presenting a nonvanishing abelian field $J_i$.
Instead, they extremize a truncation of (\ref{action4}) with $W=\Phi^I=0$.
It can easily be proven that  
the causal structure of the seed solutions is not affected by the
boosting procedure \cite{Brihaye:2005pz}.
A discussion of these solutions will be presented elsewhere 
in a more general context.

\section{Conclusions}
In this paper we have studied some properties of the nonabelian solutions in two  
versions of $D=5$, ${\cal{N}}=4$ gauged supergravity model.
In the first version, the dilaton potential have a stationary point, 
which allows for nonabelian solutions with AdS asymptotics.
We have presented numerical arguments for the existence
of both globally regular and black hole solutions.
However, we have found that the mass of these configurations generically diverges
logarithmically, the origin of this behavior residing in
both the dilaton and SU(2) sectors of the theory.
While is seems to be possible to remove the divergencies 
associated with the scalar field,
it is less clear how to define a mass for configurations with a nonvanishing gauge field.
Note that this divergence seems to be 
a generic feature of the $D=5$ nonabelian solutions,  originating in the
special scaling properties of the YM system in this spacetime dimension.
To our knowledge,
the only method to obtain $D=5$ spherically symmetric, finite mass solutions
with nonabelian fields consists in adding
corrections to the YM lagrangian consisting in higher therm of the  
YM hierarchy \cite{Brihaye:2002jg,Radu:2005mj}.

In the second part of this paper we have studied the basic properties of 
static, purely magnetic, nonabelian
solutions with unusual topology  for the case
of a Liouville dilaton potential.
Our solutions can be regarded as complements of the spherically symmetric
configurations discussed in \cite{Chamseddine:1998mc,Gubser:2001eg}.
We have added also one more member to the family of known
supersymmetric exact solution
with gravitating nonabelian fields.
However,
a nonspherical topology of the hypersurfaces $ r=const.,~t=const. $ changes drastically
the structure and properties of the solutions, leading to singular configurations.

Also, we have proposed to remove  the infrared divergencies
of the total action and mass by using a counterterm approach which
unlike background subtraction, 
does no require the identification of a reference spacetime.
It is important to note that the counterterm action proposed here gives results
that are equivalent to what one obtains using the background subtraction method.
However, we employ it because it appears to be a more general technique than
background subtraction, and it is interesting to explore the range of problems
to which it applies.

It may be instructive to give also the 
corresponding counterterm expression for the 
${\cal{N}}=4$ $D=4$  Freedman-Schwarz  gauged supergravity model.
Similar to the $D=5$ Romans model, the computation presented in the literature in this case 
makes use of a background subtraction method \cite{Gubser:2001eg}.
In this approach, the ${\cal{N}}=4$ $D=4$  
Freedman-Schwarz model with a bulk action 
 (\ref{actionFS}), presents apart from the usual Gibbons-Hawking term
 a supplementary contribution 
\begin{eqnarray}
\label{Ict-4d}
I_{ct}=
-\frac{1}{8\pi}
\int_{\partial M}d^{3}x\sqrt{-h }
(\frac{1}{2\ell}+\frac{\ell
\mathcal{R}}{4}),~~{\rm~where}~~\ell=\sqrt{-8V(\phi)}=e^{-\phi},
\end{eqnarray}
It can easily be verified that the expressions for the solutions' mass 
and action obtained by using this approach
are similar to those presented in \cite{Gubser:2001eg}.

We close  by remarking that, by using the relations in \cite{Lu:1999bw},
we can uplift all ${\cal{N}}=4$ $D=5$ configurations to ten dimensional type IIB supergravity,
with the solutions so obtained
corresponding to a five-brane wrapped in a nontrivial way.
\\
\\
{\bf Acknowledgement}
\\
This work was carried out in the framework of Enterprise--Ireland
Basic Science Research Project SC/2003/390.


\end{document}